\documentclass[runningheads]{llncs}
\usepackage{ragged2e}
\usepackage[title]{appendix}
\usepackage{xcolor}
\usepackage[T1]{fontenc}
\usepackage{longtable}
\usepackage{bm}
\usepackage{lscape}
\usepackage[figuresleft]{rotating}
\usepackage{amsmath}
\usepackage{amssymb}
\usepackage{graphicx}
\usepackage{multirow}
\usepackage{array}
\newcolumntype{C}[1]{>{\centering\let\newline\\\arraybackslash\hspace{0pt}}m{#1}}
\DeclareMathOperator*{\argmax}{\arg\!\max}

\begin{document}
\title{Hide and Seek: on the Stealthiness of Attacks against Deep Learning Systems}

\titlerunning{On the Stealthiness of Attacks against Deep Learning Systems}

\author{Zeyan Liu\inst{1} \and
Fengjun Li\inst{1} \and
Jingqiang Lin\inst{2} \and
Zhu Li \inst{3} \and
Bo Luo\inst{1}}
\institute{EECS/I2S, University of Kansas, Lawrence, KS, USA 
\email{\{zyliu, fli, bluo\}@ku.edu}
\and 
University of Science and Technology of China, Hefei, China
\email{linjq@ustc.edu.cn}
\and
University of Missouri--Kansas City, Kansas City, MO, USA
\email{lizhu@umkc.edu}}
\maketitle              
\begin{abstract}

With the growing popularity of artificial intelligence (AI) and machine learning (ML), a wide spectrum of attacks against deep learning (DL) models have been proposed in the literature. Both the evasion attacks and the poisoning attacks attempt to utilize adversarially altered samples to fool the victim model to misclassify the adversarial sample. While such attacks claim to be or are expected to be stealthy, i.e., imperceptible to human eyes, such claims are rarely evaluated. In this paper, we present the first large-scale study on the stealthiness of adversarial samples used in the attacks against deep learning. We have implemented 20 representative adversarial ML attacks on six popular benchmarking datasets. We evaluate the stealthiness of the attack samples using two complementary approaches: (1) a numerical study that adopts 24 metrics for image similarity or quality assessment; and (2) a user study of 3 sets of questionnaires that has collected 30,000+ annotations from 1,500+ responses. Our results show that the majority of the existing attacks introduce non-negligible perturbations that are not stealthy to human eyes. We further analyze the factors that contribute to attack stealthiness. We examine the correlation between the numerical analysis and the user studies, and demonstrate that some image quality metrics may provide useful guidance in attack designs, while there is still a significant gap between assessed image quality and visual stealthiness of attacks. 

\end{abstract}

\section{Introduction}

In the past decade, machine learning, especially deep learning (DL), has gained incredible success in a wide range of applications, fueling advances in every field related to big data analysis, such as computer vision, data
mining, and natural language processing. 
With the growing popularity and adoption of DL, a wide spectrum of attacks have been proposed. In particular, attacks against the \textit{integrity} of deep learning models could be roughly grouped into two categories: \textit{evasion attacks} and \textit{backdoor attacks}. In the evasion attacks, supposedly imperceptible adversarial perturbations are added to the attack samples, so that the victim models would make highly confident but erroneous classifications for these samples. In the backdoor attacks, the victim DNNs are compromised through poisoning or Trojaning, so that they ``remember'' the specially-crafted triggers (e.g., patches of pixels, shadows, or stealthy noises) as external features and classify the trigger-embedded images into wrong labels, e.g, to mis-recognize a stop sign with a yellow sticker on it as a ``go straight'' sign. 
The backdoor attacks could be further categorized into data poisoning backdoors and neural Trojans. The data poisoning backdoors inject digitally altered and mislabeled samples into the training dataset so that a malicious functionality is ``learned'' by the victim model. Meanwhile, neural Trojans alter the structure of the victim DNN by injecting a malicious sub-network that only responds to the adversarial triggers in the testing samples, so that the original task remains (mostly) unaffected. 

The majority of the existing evasion and backdoor attacks are supposed to be or claim to be \textit{stealthy}, i.e., the adversarial perturbations and the attack triggers are expected to be hardly noticeable to human eyes, so that the attacks are unlikely to be identified even if the administrators manually examine the training/testing data. However, the quality of the adversarial samples has not been carefully examined in the literature, while the stealthiness claims are rarely measured in the attack papers. To the best of our knowledge, very few existing attacks employed human evaluators to assess the stealthiness of the adversarial modifications to the attack samples \cite{nguyen2020wanet,doan2021lira} or used numerical measurements to assess the similarity between the attack and the benign images \cite{b25,li2020invisible,nguyen2020wanet}. 

In this paper, we are motivated by the questions: for the machine learning attacks proposed in the literature, how \textit{stealthy} are they? How can we \textit{quantitatively assess the attack stealthiness}? In particular, we aim to measure the \textit{numerical stealthiness} and the \textit{user-perceived stealthiness}. We have implemented 20 evasion and backdoor attacks over six popular benchmarking datasets. In \textit{numerical analysis}, we adopt 24 metrics from the literature to assess the similarities between the attack images and the corresponding benign images, or to assess the visual quality of the attack images. Some of these metrics are supposed to reflect the human visual systems or human perceptions of digital images. For \textit{user-perceived stealthiness}, we present a large-scale user study with 3 sets of questionnaires of 1,500+ responses and 30,000+ annotated images, which aim to answer the questions: could users notice the differences between the original and the adversarial samples in different attacks? If a knowledgeable or novice user is presented with attack images without the corresponding benign image, could the user identify the malicious images? We further correlate the numerical assessments with the user-perceived stealthiness and discuss our findings. 

Our contributions are: (1) We identify a largely neglected issue in adversarial ML that, while the attack samples are supposed to be or claimed to be stealthy, such features are rarely evaluated in the proposed attacks. (2) We present the first large-scale assessment and comparative study to evaluate the stealthiness of the attack samples through numerical analysis and user study. And (3) Our findings are expected to provide a better understanding of attacks against ML models, especially on how users perceive the maliciousness of the attack samples, and how auditors could benefit from our findings.   

\noindent\textbf{Ethical Considerations.} The objective of this project is to enhance the understanding of the adversarial attacks against deep learning models. All the attacks implemented in this project are previously published in the literature. All the implementations and experiments were conducted in a lab environment. We did not attack any real-world system. The user studies presented in the paper have been reviewed and approved by the Human Research Protection Program at the University of Kansas under STUDY00148002 and STUDY00148622. 

The rest of the paper is organized as follows: we introduce the background of this project in Section \ref{sec:prelim}. We present the design and results of the numerical analysis and the user study in Sections \ref{sec:num} and \ref{sec:user}, respectively, followed by discussions in Section \ref{sec:dis}. We conclude the paper in Section \ref{sec:con}.

\vspace{-1mm}
\section{Preliminaries: Deep Learning, Attacks, and Datasets}\label{sec:prelim}

\vspace{-1mm}
\subsection{Deep Learning and Adversarial Machine Learning}\label{sec:basics}

\noindent\textbf{Deep Neural Networks (DNNs).} In this paper, we focus on the standard classification task. Without loss of generality, a DNN model is defined as $f_{\theta}: X \to \mathcal{Y} $. $f_{\theta}$ maps $d$-dimensional inputs $x \in \mathcal{X} \subset {\mathbb{R}}^{d}$ into a label space $y \in \mathcal{Y}  \subset \mathbb{N+}^{k}$ with $k$ categories. Its decision making process is defined as: $p \in \mathbb{R}^{k} := f_{\theta}(x)$, 
where $p$ is the $k$-dimensional embedding of the decision  confidence in terms of the label distribution. The final output label corresponds to the maximum element in $p$. Given an annotated training dataset $D_{t} = \{(x_{i}, y_{i}): x_{i} \in \mathcal{X}, y_{i} \in \mathcal{Y}, 1\leq i \leq k\}$, the training process is to optimize the model parameters with a loss function $\mathcal{L}$, e.g., Cross-Entropy Loss:  $\theta_{optimal} = \underset{\theta}{\arg\min} \sum_{i=1}^{k} \mathcal{L}(f_{\theta}(x_{i}), y_{i})$. 

\noindent\textbf{Attacks against Deep Learning Models.} Attacks against ML/DL models could be roughly categorized into three types \cite{chakraborty2018adversarial}: (1) \textit{evasion attacks}, (2) \textit{backdoor attacks}, and (3) \textit{exploratory attacks}. Both evasion and backdoor attacks attempt to break the integrity of ML/DL models using adversarial examples that are seemingly benign, while exploratory attacks aim to break the confidentiality of the proprietary black-box models. In this paper, we focus on evasion and backdoor attacks since the majority of such attacks, implicitly or explicitly, make the stealthiness assumption on the adversarial examples. 

\noindent\textbf{Evasion Attacks.} Also known as adversarial examples, the evasion attacks add small perturbation $\delta$ to clean sample $x_{\text{beni}}$ to generate the adversarial example $x_{\text{adv}}$. $\delta$ is crafted to satisfy two objectives: (1) the magnitude of $\delta$ is limited by a perturbation budget (e.g. restrictions on ${l}^{p}$ norm). And (2) $x_{\text{adv}}$ will mislead the model to generate a wrong output $y_{\text{adv}}$, which could be untargeted (any wrong label) or targeted (a pre-selected label). 
The attack is defined as: 
\begin{align}
y_{\text{adv}} = \underset{i}{\argmax} \,f_{\theta}(x_{\text{adv}}) \neq \underset{i}{\argmax} \,f_{\theta}(x_{\text{beni}})
\end{align}

\noindent\textbf{Backdoor Attacks.} While the evasion attacks attempt to fool a benign (but vulnerable) model, backdoor attacks inject the malicious functionality into the victim DNN during model production. The attacker designs a trigger function $T$ to transform benign examples $x_{\text{beni}}$ into malicious examples $T(x_{\text{beni}})$. The backdoored model $f'_{\theta}$ learns the trigger and associates it with the target label $y_{t}$, and classifies $T(x_{\text{beni}})$ into $y_{t}$ in the testing phase: 
\begin{align}
y_{t} = \underset{i}{\argmax} \,f'_{\theta}(T(x_{beni}))
\end{align}
With different attack models, backdoors can be injected by poisoning the training data \cite{b10,b23,b25}, manually permuting the model parameters \cite{dumford2020backdooring}, or injecting malicious sub-networks into the model (Trojaning) \cite{li2021deeppayload,tang2020embarrassingly}.

\subsection{DNN Attacks Evaluated in This Study}\label{sec:attacks}

We have implemented 20 attacks (23 different settings) from the literature, including 10 evasion attacks and 10 backdoor attacks. Unless otherwise specified, we strictly follow the settings in the original papers. We briefly summarize the attacks with a special focus on adversarial perturbations or backdoor triggers. 

\noindent\textbf{Evasion Attacks.} For evasion attacks that need to specify perturbation budgets, the default budget is set to 8 (out of 255 in each RGB channel). 

\noindent$\bullet~$\textit{FGSM.} FGSM \cite{b51} is an untargeted attack that adds perturbations along the gradient sign: $\delta = \epsilon \cdot sign(\nabla_{x_{}}\mathcal{L}(f_{\theta}(x_{}, y_{})))$, where $\epsilon$ is the perturbation budget.

\noindent$\bullet~$\textit{BIM.} BIM \cite{bim} proposed a targeted variant of FGSM by iteratively updating perturbations: $ x_{t+1}= x_{t} + \alpha \cdot sign(\nabla_{x_{t}}\mathcal{L}(f_{\theta}(x_{t}, y_{adv}))$, where $x_{t}$ denotes $x_{adv}$ at the $t^{th}$ iteration, and $\alpha$ denotes the step size each iteration.

\noindent$\bullet~$\textit{MI-FGSM.} Momentum-based iterative algorithms are proposed in \cite{mifgsm}, in which the accumulated gradient is defined as: $g_{t+1}=\mu \cdot g_{t} + \frac{\nabla_{x_{t}}\mathcal{L}(f_{\theta}(x_{t}, y_{adv}))}{ \left \| \nabla_{x_{t}}\mathcal{L}(f_{\theta}(x_{t}, y_{adv})) \right \| }_{1}$, and the iterative update is: $x_{t+1}=x_{t} + \alpha \cdot sign(g_{t+1})$.

\noindent$\bullet~$\textit{PGD.} PGD \cite{pgd} improves BIM with an initialization with random noise and more iterations. The perturbation is projected onto an ${l}^{\infty}$ ball with radius $\epsilon$.

\noindent$\bullet~$\textit{AutoPGD (APGD).} AutoPGD \cite{autopgd} extends PGD by combining momentum and alternative loss functions into an efficient ensemble of attacks. AutoPGD dynamically adjusts all the parameters.

\noindent$\bullet~$\textit{FFGSM.} FFGSM \cite{ffgsm} assumes a weaker/cheaper adversary but significantly improves the efficiency of adversarial training (defense). FFGSM can be considered as a one-step variant of PGD that is initialized with uniform perturbation.

\noindent$\bullet~$\textit{Deepfool (DF).} Deepfool \cite{deepfool} computes the minimal adversarial perturbation to push the adversarial sample towards the linearized decision boundary. The gradients are calculated as: $x_{t+1}= x_{t} + \frac{\left| f_{\theta,adv}(x_{t}) - f_{\theta,beni}(x_{t}) \right|}{{\| \nabla f_{\theta,adv}(x_{t}) - \nabla f_{\theta,beni}(x_{t}) \|}_{2}^{2}}$, where $f_{\theta,adv}$ and $f_{\theta,beni}$ denote the model outputs corresponding to the target/original label.

\noindent$\bullet~$\textit{Carlini \& Wagner (CW).} C\&W \cite{cw} finds the minimum perturbation through an optimization approach. We adopt the ${l}^{2}$ constraint in the optimization objective.

\noindent$\bullet~$\textit{Smoothfool (SF).} Smoothfool \cite{smoothfool} uses low-pass filters to generate smooth adversarial perturbations, which improves attack robustness and transferability.

\noindent$\bullet~$\textit{Semantic AE (SAE).} Instead of generating artificial perturbations at the pixel level, \cite{semantic} semantically modifies the image by converting it into HSV space, and randomly shifting the Hue and Saturation components.

\begin{table}[t]
\caption{A qualitative comparison of the backdoor attacks.}\label{tbl:compare}
\vspace{-6mm}
\setlength{\tabcolsep}{0.3em}
\begin{scriptsize}
\begin{center}
\begin{tabular}{c*{9}{c}}
\hline
\multirow{2}*{} & \multirow{2}*{Datasets} & \multirow{2}*{CL} & \multirow{2}*{Attack Type}  & \multirow{2}*{Trigger Category} & \multirow{2}*{DB} & \multicolumn{3}{C{1.6cm}}{Training Method} \\
\cline{7-9}
\multirow{2}*{}  &  \multirow{2}*{~} & \multirow{2}*{~} & \multirow{2}*{~} & \multirow{2}*{~} & \multirow{2}*{~} & SC & FA & FL \\
\hline
AD & \textcircled{b}\textcircled{c} & $\times$ & One-to-One & Global Noise & $\times$ &  $\checkmark$ & $\checkmark$ & $\times$ \\ 

BN & \textcircled{a}\textcircled{b}\textcircled{d} & $\times$ & All-to-One/All-to-All & Local Patch & $\checkmark$ & $\checkmark$ & $\checkmark$ & $\times$ \\

BL & \textcircled{e} & $\times$ & One-to-One & Global Transformation & $\checkmark$  & $\checkmark$ & $\checkmark$ & $\times$ \\ 

PA & \textcircled{e} & $\times$ & One-to-One & Local Patch & $\checkmark$  & $\checkmark$ & $\checkmark$ & $\times$ \\ 

HT & \textcircled{c}\textcircled{e} & $\checkmark$ & One-to-One/All-to-One & Global Noise & $\times$  & $\times$ & $\times$ & $\checkmark$ \\ 

INS & \textcircled{b}\textcircled{c}\textcircled{e} & $\times$ & All-to-One & Global Transformation & $\times$  & $\checkmark$ & $\checkmark$ & $\times$ \\ 

IP-S & \textcircled{a}\textcircled{b}\textcircled{c} & $\times$ & One-to-One & Global Transformation & $\checkmark$  & $\checkmark$ & $\checkmark$ & $\times$\\ 

IP-A & \textcircled{a}\textcircled{b}\textcircled{c} & $\times$ & One-to-One & Global Noise & $\checkmark$ & $\checkmark$ & $\checkmark$ & $\times$\\ 

RP & \textcircled{a}\textcircled{b} & $\checkmark$ & All-to-One & Global Transformation & $\times$ & $\checkmark$ & $\checkmark$ & $\times$ \\ 


WM & \textcircled{g}\textcircled{h} & $-$ &  All-to-One & Local Patch & $\checkmark$ & $\times$ & $\times$ & $\checkmark$ \\ 

SQ & \textcircled{g}\textcircled{h} & $-$ &  All-to-One & Local Patch & $\checkmark$ & $\times$ & $\times$ & $\checkmark$ \\ 

TNet & \textcircled{b}\textcircled{e}\textcircled{f}\textcircled{j} & $-$ & All-to-One/All-to-All & Local Patch & $\checkmark$  & - & - & - \\ 

WN & \textcircled{a}\textcircled{b}\textcircled{c}\textcircled{i} & $\times$ & All-to-One & Global Transformation & $\times$  & $\checkmark$ & $\checkmark$ & $\times$ \\

\end{tabular}
\end{center}
\begin{justify}
\vspace{-1mm}
\noindent\textbf{Datasets in the published paper: }\textcircled{a} MNIST. \textcircled{b} GTSRB. \textcircled{c} CIFAR10. \textcircled{d} US \& Swedish Traffic Signs.  \textcircled{e} ImageNet. \textcircled{f} YouTube Aligned Face.  \textcircled{g} VGG Face.  \textcircled{h} LFW. \textcircled{i} CelebA. \textcircled{j} Pubfig. 

\noindent \textbf{CL}: Clean label poisoning. \textbf{DB}: Whether the benign dataset is unavailable. \textbf{SC}: Train the model from scratch. \textbf{FA}: Finetune all the layers. \textbf{FL}: Finetune top layers with other layers frozen.

\end{justify}
\end{scriptsize}
\vspace{-6mm}
\end{table}

\noindent\textbf{Backdoor Attacks.} A detailed comparison of the backdoor attack models and settings is shown in Table~\ref{tbl:compare}. 

\noindent$\bullet~$\textit{Advdoor (AD).} Advdoor \cite{zhang2021advdoor} utilizes a Targeted Universal Adversarial Perturbation (TUAP) trigger, which is claimed to be ``small magnitude and human imperceptible''. Input-specific perturbations are iteratively aggregated to generate UAP, where existing evasion attacks (e.g., Deepfool) could be employed.

\noindent$\bullet~$\textit{Badnets (BN).} 
Badnets \cite{b23} patch a unique black-and-white square on the corner of training samples. We follow the settings in \cite{b23} and use a $4{\times}4$ trigger. 

\noindent$\bullet~$\textit{Blended \& Physical Accessory (BL \& PA).} \cite{b10} employs two trigger generation strategies: blended injection (BL) for simple digital attacks and physical accessory injection (PA) for real-world physical attacks. We follow \cite{b10} to use the hello kitty image in BL, and follow \cite{zeng2021rethinking} to use mug patches in PA.

\noindent$\bullet~$\textit{HiddenTrigger (HT).} 
HT \cite{b34} is a clean-label backdoor, which claims the poisoned samples to be ``natural with correct labels'', while the triggers are hidden until test time. In HT, we assess the stealthiness of poisoned training samples. 

\noindent$\bullet~$\textit{Instagram (INS).} \cite{liu2019abs} proposed feature-space attacks that generate triggers with Instagram filters Nashville and Gotham. We implement the Nashville filter.

\noindent$\bullet~$\textit{Invisible Perturbation (IP).} \cite{b25} propose two strategies to generate ``hardly perceptible'' backdoor triggers: patterned static perturbation masks (IP-S) and targeted adaptive perturbation masks (IP-A). 

\noindent$\bullet~$\textit{Ramp Signal (RP).} \cite{b59} uses ramp signals as triggers in the poisoned samples.

\noindent$\bullet~$\textit{TrojanNN (TN).} \cite{b13} use reverse-engineered adversarial triggers to maximize the activation of anchor neurons in the DNN. 
Among the trigger patterns in \cite{b13}, we have implemented Watermark (WM) and Square (SQ) in this study.  

\noindent$\bullet~$\textit{TrojanNet (TNet).} \cite{tang2020embarrassingly} injects into the victim DNN a small neural component, which is activated by a predefined trigger pattern to flip the DNN output. We follow the settings in \cite{tang2020embarrassingly} to implement a $4\times4$ black-and-white square trigger.

\noindent$\bullet~$\textit{WaNet (WN).} \cite{nguyen2020wanet} uses elastic image warping as triggers. Human inspection experiments in \cite{nguyen2020wanet} demonstrate its outstanding stealthiness.

\vspace{1mm}
\noindent\textbf{Attack categorization.} Based on the properties of the injected adversarial perturbations, we classify the attacks into three categories: (1) Gradient-based noise (or global noise): FGSM, BIM, PGD, MI-FGSM, DF, FFGSM, APGD, CW, SF, AD, HT, IP-A. They make small modifications to a large number of pixels to achieve a global optimization objective. They attack the unrobust deep features in the victim DNN, while the main contents (the semantic meanings) of the adversarial images remain unchanged. (2) Local patches: BN, PA, WM, SQ, TNet. They are all backdoor attacks that modify a small region on adversarial images to build salient visual features. The triggers are robust, but they are usually detectable by trigger reconstruction defenses, e.g., \cite{b36,b32}. (3) Global transformation: SAE, BL, INS, IP-S, RP, WN. They make global changes to the victim images, which may significantly change the appearances of the images.

\begin{figure}[t]
\centering
	\includegraphics[width=0.98\linewidth]{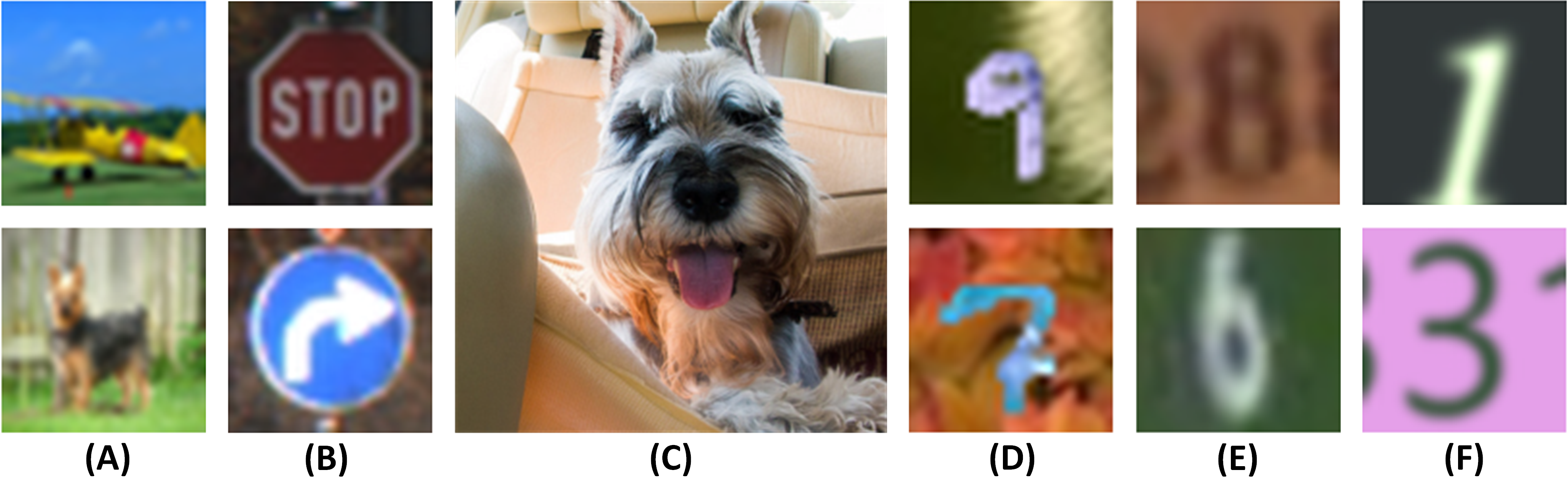}\vspace{-2mm}
	\caption{Sample images from the benchmarking datasets used in this study.\vspace{-2mm}}
	\label{fig:dataset}
\end{figure}

\subsection{The Datasets}\label{sec:dataset}
In this study, we adopt six datasets that are frequently used as benchmarks in both the machine learning community and the adversarial ML community. Please note that, while most of the attacks were originally implemented on a subset of these datasets, we re-implemented all of them on all six datasets.


\vspace{1mm}
\noindent\textbf{A. CIFAR-10.} The CIFAR dataset \cite{krizhevsky2009learning} contains 60K tiny images (32$\times$32) labeled in 10 classes, such as automobile, bird, cat, ship, truck, etc. 

\noindent\textbf{B. GTSRB.} The German Traffic Sign Recognition Benchmark dataset \cite{Houben-IJCNN-2013} has 51K images of real-world traffic signs in 43 classes. Images are scaled to 32$\times$32.

\noindent\textbf{C. ImageNet.} ImageNet is a large-scale image classification and object recognition benchmark dataset. The dataset used in the ImageNet Large Scale Visual Recognition Challenge (ILSVRC) 2012-14 contains 1.35 million images that are manually labeled into 1,000 classes \cite{ILSVRC15}. The images are usually cropped to 224$\times$224. They are significantly larger than the small images in other datasets. 

\noindent\textbf{D. MNIST-M.} The MNIST dataset\footnote{Available at: \url{http://yann.lecun.com/exdb/mnist/}} contains 70,000 binary images of hand-written digits. The MNIST-M dataset adds color to MNIST by extracting patches from the BSDS500 dataset and combining them with MNIST \cite{ganin2016domain}. 

\noindent\textbf{E. SVHN.} The Street View House Numbers (SVHN) is another digit classification benchmark dataset, which contains 600,000 digit images that are extracted from house numbers in Google Street View \cite{netzer2011reading}.

\noindent\textbf{F. SYNDIGIT.} The Synthetic Digits dataset contains 12,000 images that are synthetically generated by placing digits of various fonts, colors, and directions over random backgrounds \cite{ganin2015unsupervised}. The MNIST-M, SVHN, and SYNDIGIT datasets are often used in transfer learning and attacks against transfer learning. 

Samples of benign images from the datasets are shown in Figure~\ref{fig:dataset}. The corresponding attack images are generated by exploiting the attacks introduced in Section~\ref{sec:attacks} on images sampled from the six datasets. They may contain noise (a.k.a. adversarial perturbation) that is supposed to be invisible to human eyes, or adversarial patterns such as a sticky note on a stop sign.

\begin{figure}[t]
\centering
	\includegraphics[width=0.98\linewidth]{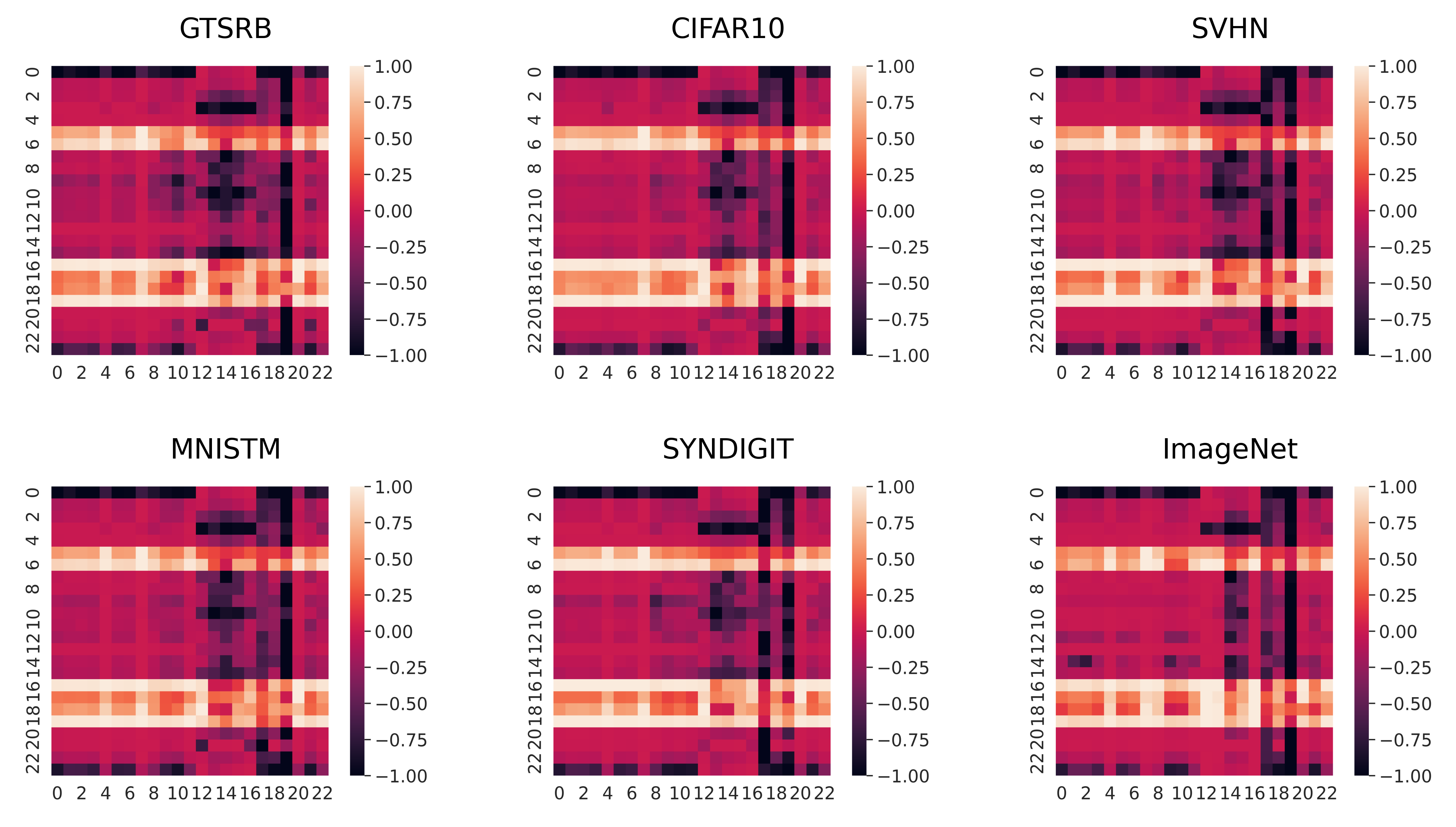}\vspace{-2mm}
	\caption{Normalized numerical assessment results. X-axis: attacks. Y-axis: metrics. \vspace{-2mm}}
	\label{fig:heatmap}
\end{figure}

\section{The Numerical Analysis}\label{sec:num}

\subsection{Metrics and Basic Statistics}\label{sec:metric}
In this paper, we adopt three categories of metrics in the numerical analysis. 

\noindent\bm{${l}^{p}$} \textbf{Norms}: ${l}^{p}$ norms are widely used in adversarial example attacks to limit the magnitude of the perturbation. In the rest of the paper, we always use ${l}^{p}$ norms normalized by image size, i.e., 
 ${l}^{p}(x_{\text{adv}}-x_{\text{beni}})/|x_{\text{beni}}|$, where $p=\{0, 1, 2, \infty\}$. 

\noindent\textbf{Basic Metrics}: Most popular metrics in the literature for image similarities: MSE (mean square error), PSNR (peak signal-to-noise ratio), and SSIM \cite{SSIM}.

\noindent\textbf{Additional Metrics}: To further explore how the advanced image similarity or quality measurements may fit into this problem, we have implemented 17 additional metrics. They are roughly categorized into 5 groups:  

\vspace{1mm}
\noindent\textbf{(1)} Simple spatial domain metrics: Average Hash (aHash) \cite{fei2015visual}, Difference Hash (dHash) \cite{fei2015visual} and variants of ${l}^{p}$ norms: relative error in ${l}^{1}$ (RE), Elastic-Net regularization \cite{Elastic}, ${l}^{1}$ with cosine similarity \cite{l1cosine} and ${l}^{1}$ with clip function \cite{l1clip}.

\noindent\textbf{(2)} Structural similarity/quality metrics: Universal Image Quality index (UQI) \cite{UQI} and Multi-scale SSIM (MS-SSIM) \cite{MSSSIM}. 

\noindent\textbf{(3)} Transform domain metrics: Perceptual Hash (pHash) \cite{fei2015visual}, Wavelet Hash (wHash), Visual Information Fidelity (VIF) \cite{VIF}, Feature Similarity index (FSIM) \cite{FSIM}, ${l}^{1}$ loss in the frequency domain (FDL) and focal frequency loss (FFL) \cite{jiang2021focal}.

\noindent\textbf{(4)} Spectral similarity: ERGAS \cite{ERGAS} and SAM\cite{SAM}.

\noindent\textbf{(5)} Metrics based on deep learning: LPIPS \cite{LPIPS}.

{
\setlength{\tabcolsep}{0.2em}
\begin{table}[t]
\caption{Image similarity/quality measurements for evasion attacks. Blue: the best attack (most similar or highest quality) for each metric; red: the worst attack.}\label{tbl:evasion}
\vspace{-1mm}
\begin{scriptsize}
\begin{center}
\begin{tabular}{ccc|ccccccccc}
\hline

\bfseries{Metrics} & {Ideal} & {Better} & FGSM & BIM & PGD & MIFGSM & DF & FFGSM & APGD & CW & SF
\\\hline

${l}^{0}$  &  0  &  small  &  0.991  &  0.863  &  0.971  &  0.992  &  0.709  &  \textcolor{red}{0.994}  &  0.971  &  \textcolor{blue}{0.647}  &  0.829 \\\hline

${l}^{1}$  &  0  &  small  &  \textcolor{red}{7.632}  &  5.189  &  5.320  &  6.282  &  2.286  &  6.630  &  6.068  &  \textcolor{blue}{1.153}  &  4.484 \\\hline

${l}^{2}$  &  0  &  small  &  \textcolor{red}{0.130}  &  0.101  &  0.101  &  0.114  &  0.060  &  0.116  &  0.112  &  \textcolor{blue}{0.032}  &  0.121 \\\hline

${l}^{\infty}$  &  0  &  small  &  \textcolor{blue}{0.003}  &  \textcolor{blue}{0.003}  &  \textcolor{blue}{0.003}  &  \textcolor{blue}{0.003}  &  0.007  &  \textcolor{blue}{0.003}  &  \textcolor{blue}{0.003}  &  0.004  &  {0.012} \\\hline

MSE  &  0  &  small  &  59.20  &  34.96  &  35.07  &  45.09  &  24.55  &  47.48  &  42.51  &  \textcolor{blue}{3.686}  &  \textcolor{red}{65.26} \\\hline

PSNR  &  $\infty$  &  large  &  \textcolor{red}{30.42}  &  32.76  &  32.75  &  31.61  &  40.98  &  31.37  &  31.92  &  \textcolor{blue}{44.32}  &  34.40 \\\hline

SSIM  &  1  &  large  &  \textcolor{red}{0.928}  &  0.957  &  0.957  &  0.943  &  0.984  &  0.940  &  0.953  &  \textcolor{blue}{0.996}  &  0.963 \\\hline

LPIPS  &  0  &  small  & \textcolor{red}{0.011}  &  0.005  &  0.005  &  0.007  &  0.001  &  0.008  &  0.006  &  \textcolor{blue}{1.9e-4}  &  0.003 \\\hline

aHash  &  0  &  small  &  0.642  &  0.625  &  0.533  &  0.670  &  0.405  &  0.587  &  0.660  &  \textcolor{blue}{0.320}  &  \textcolor{red}{1.442} \\\hline

dHash  &  0  &  small  &  3.035  &  2.727  &  2.642  &  2.908  &  1.633  &  2.663  &  2.973  &  \textcolor{blue}{1.245}  &  \textcolor{red}{4.508} \\\hline

pHash  &  0  &  small  &  1.813  &  1.620  &  1.590  &  1.763  &  0.800  &  1.707  &  1.700  &  \textcolor{blue}{0.610}  &  \textcolor{red}{3.780} \\\hline

wHash  &  0  &  small  &  1.613  &  1.400  &  1.318  &  1.477  &  0.658  &  1.473  &  1.438  &  \textcolor{blue}{0.520}  &  \textcolor{red}{2.527} \\\hline

FDL  &  0  &  small  &   \textcolor{red}{46.28}  &  34.56  &  34.64  &  39.96  &  15.65  &  41.51  &  \textcolor{blue}{7.49}  &  37.73  &  21.62 \\\hline

FFL  &  0  &  small  &  35.27  &  20.84  &  20.92  &  26.76  &  15.11  &  28.32  &  25.31  &  \textcolor{blue}{2.233}  &  \textcolor{red}{46.95} \\\hline

ERGAS  &  0  &  small  &  5108  &  8340  &  \textcolor{red}{1.0e4}  &  5319  &  1971  &  5192  &  4282  &  \textcolor{blue}{1153}  &  4786 \\\hline

SAM  &  0  &  small  &   \textcolor{red}{0.080}  &  0.060  &  0.059  &  0.067  &  0.030  &  0.070  &  0.065  &  \textcolor{blue}{0.016}  &  0.057 \\\hline

MS-SSIM  &  1  &  large  &  0.992  &  0.995  &  0.995  &  0.993  &  0.998  &  0.993  &  0.994  &  \textcolor{blue}{0.999}  &  \textcolor{red}{0.983} \\\hline

FSIM  &  1  &  large  &  \textcolor{red}{0.767}  &  0.782  &  0.783  &  0.768  &  0.888  &  0.768  &  0.775  &  \textcolor{blue}{0.920}  &  0.848 \\\hline

VIF  &  1  &  $\sim$1  &  \textcolor{red}{0.513}  &  0.587  &  0.587  &  0.551  &  0.752  &  0.543  &  0.561  &  \textcolor{blue}{0.849}  &  0.600 \\\hline

UQI  &  1  &  large  &  \textcolor{red}{0.990}  &  \textcolor{red}{0.990}  &  0.991  &  0.990  &  0.997  &  0.991  &  0.992  &  \textcolor{blue}{0.998}  &  \textcolor{red}{0.990} \\\hline

Elastic  &  0  &  small  &  \textcolor{red}{17.95}  &  11.14  &  11.27  &  14.04  &  6.74  &  14.80  &  13.36  &  \textcolor{blue}{1.659}  &  16.64 \\\hline

RE  &  0  &  small  &  \textcolor{red}{4.457}  &  2.673  &  2.509  &  3.974  &  1.575  &  3.626  &  3.088  &  \textcolor{blue}{0.038}  &  0.495 \\\hline

${l}^{1}$ Cos  &  0  &  small  &  \textcolor{red}{46.07}  &  31.35  &  32.12  &  37.96  &  13.76  &  40.02  &  36.63  &  \textcolor{blue}{6.952}  &  27.05 \\\hline

${l}^{1}$ Clip  &  0  &  small  &  \textcolor{red}{7.632}  &  5.189  &  5.320  &  6.282  &  1.919  &  6.630  &  6.068  &  \textcolor{blue}{1.146}  &  3.406 \\\hline

\end{tabular}
\end{center}
\end{scriptsize}
\vspace{-5mm}
\end{table}
}

For each attack, we have generated 1,000 adversarial images for each dataset. We employ the metrics introduced above to assess the attack images and report the mean assessment values in Tables~\ref{tbl:evasion} and \ref{tbl:backdoor}. In Table~\ref{tbl:evasion}, we provide a high-level illustration of how the numbers could be interpreted. Column ``ideal'' denotes the output of the metric on two identical images, while ``better'' denotes whether a larger or a smaller value indicates ``more similar'' images (``better'' attack). For example,  MSE generates 0 for two identical images, and a smaller value corresponds to images with higher similarity, i.e., the attack is stealthier in this metric. Meanwhile, Figure~\ref{fig:heatmap} shows the detailed distribution in each dataset after min-max normalization, in which a lighter color ($\sim 1$) indicates a stealthier attack. As shown in the figure, the measurements for each attack demonstrate very similar patterns across six datasets, i.e., the dataset is NOT a significant factor in numerical assessments of attack stealthiness.

{
\setlength{\tabcolsep}{0.2em}
\begin{table}[t]
\caption{Image similarity/quality measurements for backdoor attacks. Blue: the best attack (most similar or highest quality) for each metric; red: the worst attack.}\label{tbl:backdoor}
\vspace{-5mm}
\begin{scriptsize}
\begin{center}
\begin{tabular}{cccc|ccccc|cccccc}
\hline

\multirow{2}*{\bfseries{Metrics}}&\multicolumn{3}{C{2.1cm}|}{\bfseries{Global Noises}}&\multicolumn{5}{C{3.5cm}|}{\bfseries{Local Patches}} & \multicolumn{6}{C{4.2cm}}{\bfseries{Global Transformation}}
\\

\multirow{2}*{~} & AD & HT & IPA & BN & PA & WM & SQ & TNet & SAE & BL & INS & IPS & RP & WN
\\\hline

${l}^{0}$  &  0.937  &  0.973  &  0.946  &  \textcolor{blue}{0.015}  &  0.136  &  0.078  &  0.057  &  \textcolor{blue}{0.015}  &  0.888  &  0.983  &  \textcolor{red}{0.991}  &  0.263  &  0.878  &  0.731 \\\hline

${l}^{1}$  &  8.531  &  9.866  &  5.028  &  \textcolor{blue}{1.993}  &  7.822  &  7.884  &  5.438  &  2.102  &  35.61  &  24.50  &  \textcolor{red}{49.38}  &  2.471  &  11.81  &  3.413 \\\hline

${l}^{2}$  &  0.175  &  0.188  &  0.104  &  0.325  &  0.472  &  0.588  &  0.499  &  0.334  &  0.829  &  0.473  &  \textcolor{red}{1.037}  &  \textcolor{blue}{0.083}  &  0.226  &  0.105\\\hline

${l}^{\infty}$  &  0.006  &  0.005  &  0.004  &  0.068  &  0.057  &  \textcolor{red}{0.072}  &  0.069  &  0.069  &  0.040  &  0.020  &  0.060  &  \textcolor{blue}{0.003}  &  0.006  &  0.013\\\hline

MSE  &  113.2  &  121.0  &  48.12  &  333.0  &  751.8  &  1302  &  948.4  &  355.1  &  3231 &  944.3  &  \textcolor{red}{4225}  &  \textcolor{blue}{24.46}  &  181.8  &  45.86 \\\hline

PSNR  &  28.09  &  27.37  &  \textcolor{blue}{34.59}  &  25.21  &  22.39  &  17.40  &  19.01  &  24.97  &  16.51 &  19.78  &  \textcolor{red}{12.57}  &  34.25  &  25.55  &  33.61\\\hline

SSIM  &  0.864  & 0.868  &  0.962  &  0.953  &  0.784  &  0.660  &  0.874  &  0.898  &  \textcolor{red}{0.646} &  0.898  &  0.722  & 0.973  &  0.873  &  \textcolor{blue}{0.983}\\\hline

LPIPS  &  0.019  &  0.022  &  0.007  &  0.034  &  0.037  &  \textcolor{red}{0.122}  &  0.072  &  0.021  &  0.066 &  0.007  &  0.083  &  \textcolor{blue}{8.7e-4}  &  0.021  &  0.003\\\hline

aHash  &  0.972  &  1.376  &  0.553  &  1.670  &  6.810  &  6.461  &  6.288  &  1.842  &  5.170 &  2.003  &  \textcolor{red}{11.71}  &  \textcolor{blue}{0.362}  &  0.610  &  1.157\\\hline

dHash  &  3.927  &  4.664  &  2.777  &  2.453  &  6.782  &  8.148  &  4.372  &  2.837  &  6.350  &  5.158  &  \textcolor{red}{11.83}  &  \textcolor{blue}{1.223}  &  3.473  &  2.708\\\hline

pHash  &  2.510  &  3.275  &  1.620  &  7.747  &  12.85  &  10.93  &  12.95  &  7.993  &  6.383 &  3.897  &  \textcolor{red}{11.21}  &  \textcolor{blue}{0.360}  &  1.513  &  2.660\\\hline

wHash  &  2.248  &  2.688  &  1.487  &  1.687  &  7.028  &  7.662  &  6.258  &  1.873  &  5.545 &  3.358  &  \textcolor{red}{13.36}  &  \textcolor{blue}{0.437}  &  4.465  &  1.725\\\hline

FDL  &  58.69  &  64.04  &  29.68  &  14.00  &  43.56  &  136.1  &  60.64  &  18.25  &  171.8 &  68.57  &  \textcolor{red}{220.6}  &  \textcolor{blue}{12.14}  &  36.02  &  23.11\\\hline

FFL  &  65.98  &  71.36  &  28.38  &  283.3  &  546.8  &  657.3  &  525.4  &  278.0  &  2382 &  806.7  &  \textcolor{red}{3195}  &  \textcolor{blue}{24.24}  &  179.2  &  32.32\\\hline

ERGAS  &  1.3e4  &  8865  &  6329  &  \textcolor{blue}{2549}  &  1.0e4  &  2.7e4  &  1.2e4  &  5670  &  2.0e4 &  1.5e4  &  \textcolor{red}{4.3e4}  &  4986  &  1.0e4  &  3070\\\hline

SAM  &  0.110  & 0.125  &  0.060  &  0.171  &  0.225  &  0.313  &  0.288  &  0.177  &  0.289 &  0.082  &  \textcolor{red}{0.381}  &  \textcolor{blue}{0.045}  &  0.135  &  0.053\\\hline

MS-SSIM  &  0.980  &  0.982  &  0.996  &  0.985  &  \textcolor{red}{0.818}  &  0.842  &  0.874  &  0.957  &  0.805 &  0.952  &  0.881  &  \textcolor{blue}{0.999}  &  0.966  &  0.996\\\hline

FSIM  &  0.734  &  0.696  &  0.790  &  0.960  &  0.813  &  0.794  &  0.840  &  0.948  &  0.704  &  0.821  &  \textcolor{red}{0.616}  &  \textcolor{blue}{0.992}  &  0.747  &  0.870\\\hline

VIF  &  0.393  &  0.410  &  0.626  &  \textcolor{blue}{0.879}  &  0.433  &  \textcolor{red}{0.264}  &  0.679  &  0.697  &  0.363 &  0.587  &  0.528  &  0.670  &  0.405  &  0.619\\\hline

UQI  &  0.979  &  0.979  &  0.991  &  0.988  &  0.948  &  0.899  &  0.959  &  0.972  &  \textcolor{red}{0.781}  &  0.939  &  0.799  &  \textcolor{blue}{0.992}  &  0.969  &  0.996\\\hline

Elastic  &  29.47  &  32.09  &  13.65  &  68.20  &  156.6  &  266.6  &  194.0  &  72.70  &  674.6 &  208.5  &  \textcolor{red}{884.5}  &  \textcolor{blue}{6.869}  &  45.80  &  11.90\\\hline

RE  &  7.872  &  7.966  &  4.430  &  80.77  &  0.114  &  0.065  &  0.065  &  50.28  &  \textcolor{red}{256.6} &  0.266  &  153.8  &  0.072  &  21.41  &  \textcolor{blue}{0.064}\\\hline

${l}^{1}$ Cosine  &  51.58  &  59.55  &  30.38  &  \textcolor{blue}{12.19}  &  47.79  &  49.05  &  33.75  &  12.83  &  \textcolor{red}{215.5} &  147.2  &  297.8  &  14.91  &  71.21  &  20.58\\\hline

${l}^{1}$ Clip  &  6.406  &  7.857  &  4.213  &  0.153  &  1.138  &  0.608  &  0.393  &  \textcolor{blue}{0.153}  &  7.356 &  8.466  &  \textcolor{red}{9.307}  &  2.467  &  7.896  &  2.682 \\\hline

\end{tabular}
\end{center}
\end{scriptsize}
\vspace{-5mm}
\end{table}
}

\subsection{Numerical Analysis of Attack Stealthiness}
\noindent\textbf{${l}^{p}$ Norms (normalized by image size) and attack categorization. } The ${l}^{0}$ norm indicates the portion of pixels that are modified, while the ${l}^{\infty}$ norm indicates the maximum change made to a pixel. Both ${l}^{1}$ and ${l}^{2}$ denote the overall amplitude of the added adversarial perturbations, while ${l}^{2}$ is more sensitive to large changes. 
As shown in Tables~\ref{tbl:evasion} and~\ref{tbl:backdoor}, the $l^p$ norms are good indicators of the types of the added perturbations: (1) In general, global noise attacks always produce large $l^0$ (close to 1) but moderate $l^1$, $l^2$, and smallest $l^\infty$ (mean=0.0044). By design, they introduce a small change (often explicitly limited by $l^p$ norm) to each pixel. (2) Local patches generate smallest $l^0$ (mean=0.06) and largest $l^\infty$ (mean=0.067). The mean $l^1$ is smaller than noise-based attacks', while the mean $l^2$ is larger. Although local patches are small in size, they make significant modifications to the pixel values in order to build strong visual features, which significantly impact the $l^2$ and $l^\infty$ norms. (3) Global transformation attacks introduce moderate changes globally, which result larger $l^0$, $l^1$, $l^2$, but moderate $l^\infty$ (mean=0.024). IP-S is an exception in that it employs a scattered global pattern, so that it moderately modifies a relatively small portion of pixels ($l^0=0.263$, $l^1=2.471$) distributed over the entire image. 

\noindent\textbf{MSE, PNSR, and SSIM.} CW produces the best overall MSE (3.686) and PNSR (44.32). Gradient-based attacks all result in smaller MSE ([3.686, 121]), local patch backdoors produce moderate MSE ([333, 1302]), while global transformation attacks generate variable MSE (three attacks in [24.46, 181.8] and others in [944.3, 4225]). PNSR generates very similar rankings and patterns as MSE. SSIM imitates the Human Visual System (HVS) to measure the structural similarity based on luminance, contrast, and structure. CW again produces the best SSIM (0.996). Among local patch backdoors, BN generates the best SSIM of 0.953, while WM gives the worst SSIM of 0.660. Among global transformation attacks, WN provides the best SSIM of 0.983, while SAE provides the worst SSIM of 0.646. In our experiments, both PNSR and SSIM are highly (negatively) correlated with MSE, with Spearman correlation coefficients of -0.92 and -0.86. 

In summary, noise-based attacks significantly outperform the other two categories in all three metrics with averages of 53.35, 33.38, and, 0.943 respectively. Two global transformation attacks, SAE and INS, perform the worst in all three metrics, which is understandable since they have very different design philosophies that do not seek to minimize perturbation. 

\noindent\textbf{Additional Image Quality Metrics. } IP-S appears to be the most stealthy with dHash, pHash, wHash, and FSIM. BN performs the best with VIF and ${l}^{1}$, while CW is the best for all other measurements. In general, gradient noise attacks perform better than the other two categories in most cases. IP-S again appears to be an outlier among global transformation attacks. Most of the metrics demonstrate strong correlations with at least one of MSE, PNSR, or SSIM (absolute correlation coefficient > 0.8). These metrics will be further explored in comparison with the results from the user study.

\section{The User Study}\label{sec:user}

In this section, we evaluate the likelihood of adversarial samples escaping human users. We mimic an auditing scenario, in which auditors/defenders examine potentially suspicious images in an attempt to identify true attack samples.

\subsection{Research Design}

Based on the level of knowledge the auditor/defender has about the dataset, we define three defense models:

\noindent\textbf{Model I. Informed Defenders (White Box)}. The Informed Defenders have full knowledge of the target dataset. They are capable of making side-by-side comparisons of the original and the adversarial samples in a white-box manner. They represent the strongest type of defenders, who more likely exist as the owners of the original datasets in defense against training data poisoning. 

\noindent\textbf{Model II. Knowledgeable Defenders (Grey Box)}. The Knowledgeable Defenders have reasonable knowledge about the benign dataset. When they manually inspect the potentially adversarial samples, they can refer to benign samples from the same dataset. However, they do not possess the exact original images that correspond to the suspect samples, i.e., no side-by-side comparison. 

\noindent\textbf{Model III. Referenceless Defenders (Black Box)}. The Referenceless Defenders have little or no access to the original dataset. They inspect the potentially adversarial samples without referring to benign samples. They represent the weakest type of defenders, who often exist among the novice users, who download and reuse pre-trained models from the Internet, e.g., Model Zoo. 

In Defender Model I, attack stealthiness is defined as the \textit{invisibility} of the adversarial perturbations--when the auditor notices the noise/patch, the attack is likely discovered. In Models II and III, attack stealthiness is defined as \textit{visual benignness}, i.e., the attack examples are expected to be visually consistent with the examiners' psychophysical perceptions of normal images. The perceptions will differ with or without knowing the dataset.

To mimic each type of defenders, we have designed three user studies\footnote{The user studies have been approved by the Human Research Protection Program at the University of Kansas under STUDY00148002 and STUDY00148622.}. 
In each user study, an IRB information statement is first displayed to the participant, followed by a link to the questionnaire. In each questionnaire, 20 (sets of) images are presented to the participant for inspection. ImageNet images are downsized to 200$\times$200, while the smaller images (e.g., GTSRB, CIFAR) are enlarged to 4$\times$ the original size, and padded with white padding to 200$\times$200. 

\begin{figure}[t]
\centering
	\includegraphics[width=0.98\linewidth]{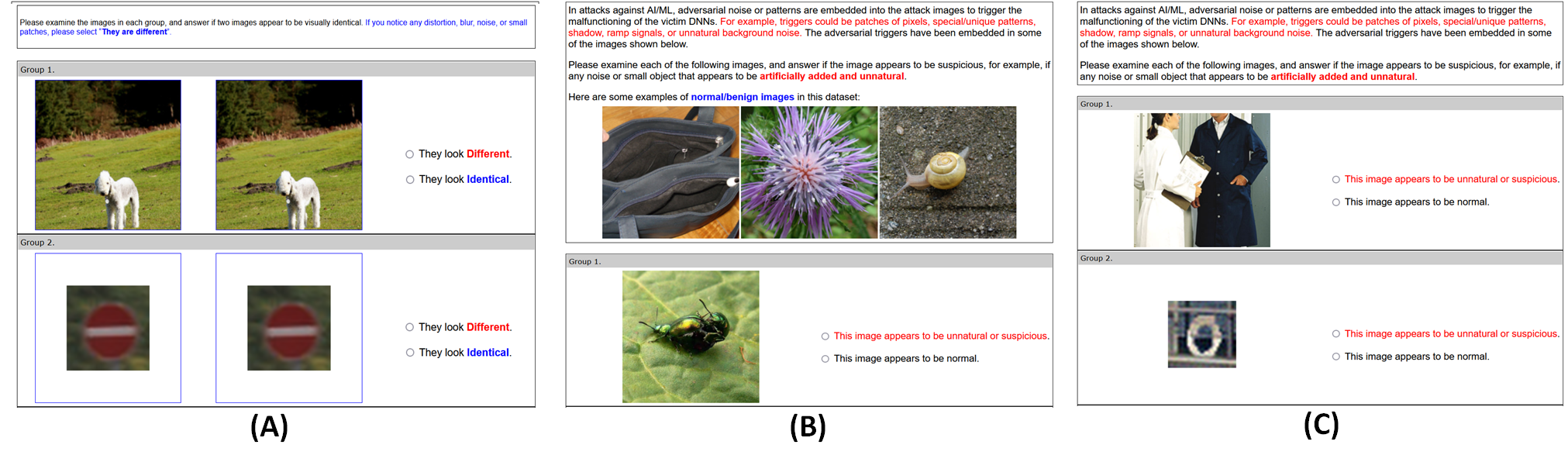}\vspace{-2mm}
	\caption{Sample questionnaires used in the user study.\vspace{-5mm}}
	\label{fig:survey}
\end{figure}

\noindent\textbf{Exp I. Informed Defenders (White Box).} The questionnaire contains 20 pairs of images: the first in each pair is a benign image randomly selected from the datasets introduced in Section~\ref{sec:dataset}; the second is the adversarial image that is altered by a random attack in Section~\ref{sec:attacks}. For quality control, we include 10\% benign images as the second, i.e., two images are numerically identical. Part of the questionnaire is shown in Figure~\ref{fig:survey}~(A). We ask the user to inspect each pair and identify whether the two images look identical. 

\noindent\textbf{Exp II. Knowledgeable Defenders (Grey Box).} As shown in Figure~\ref{fig:survey} (B), the questionnaire first displays three benign images from a random dataset to be used by the participant as reference images. The questionnaire then shows 20 different images from \textit{the same dataset} as the reference images. For each image, we ask the participant to select if it appears to be benign or suspicious. 

\noindent\textbf{Exp III. Referenceless Defenders (Black Box).} As shown in Figure~\ref{fig:survey} (C), the questionnaire is very similar to the Knowledgeable Defenders' questionnaire, except that no reference image is displayed. 

\subsection{The User-Perceived Attack Stealthiness}\label{sec:res}
We sent the questionnaires to senior/graduate CS students in three institutions which the authors are affiliated with. We intentionally select CS students because they have basic understanding of computer systems and programming, while some of them have prior knowledge of AI/ML or even adversarial ML. They better mimic the system administrators or ML/AI users in real world applications. In six weeks, we collected 1,526 responses with 30,369 total annotations, including 504, 512, and 510 responses, with 9,991, 10,209, and 10,169 annotations to Experiments 1, 2 and 3, respectively. 

\begin{figure}[t]
\centering
	\includegraphics[width=0.98\linewidth]{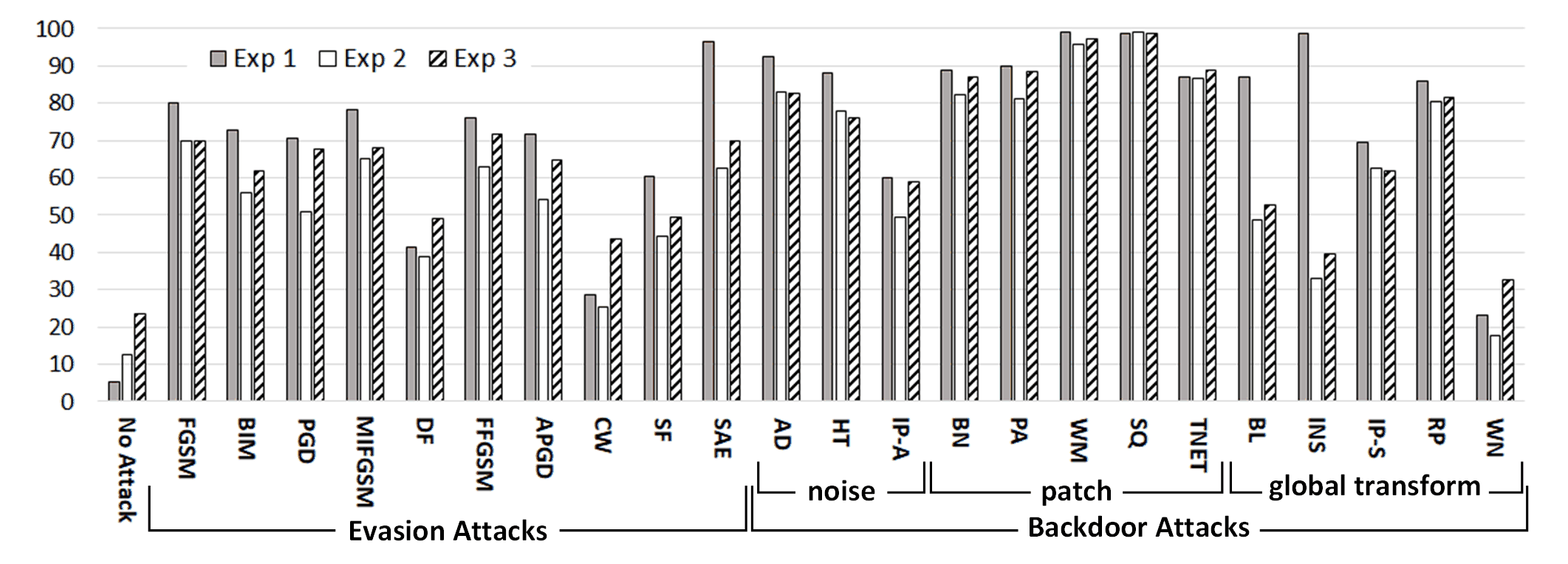}\vspace{-4mm}
	\caption{Detection rates (\%) from user studies. Exp I: informed defender; II: knowledgeable defender;  III:  referenceless defender. A lower rate means a stealthier attack. \vspace{-4mm}}
	\label{fig:bar}
\end{figure}

We present the \textit{detection rates} from all the experiments in Fig.~\ref{fig:bar}. The detection rate is defined as the proportion of annotations that labeled a pair of images as ``different'' (Exp 1) or labeled an image as ``unnatural or malicious'' (Exp 2 and 3). A lower detection rate indicates a stealthier attack. 

\vspace{1mm}
\noindent\textbf{Exp I. Informed Defenders.} The average detection rate is 75.9\%. The detection rates of 17 attacks were higher than 70\%, while 11 attacks were detected in more than 85\% of the tests. The statistics show that majority of the adversarial samples are highly noticeable, i.e., not stealthy, in side-by-side comparisons with the corresponding benign images. The median task completion time was 91 seconds (for 20 image pairs), indicating the level of effort needed in an audit of the training or testing dataset. WN and CW significantly outperform the other attacks, with detection rates at 28.5\% and 23.3\%, while detection rates of SQ and WM are as high as 98.9\% and 98.7\%. In general, noise-based evasion attacks appear to be stealthier than the others with a 62.0\% mean detection rate, while patch-based backdoors are the least stealthy with a 92.7\% mean detection rate. 

\vspace{1mm}\noindent\textbf{Exp II. Knowledgeable Defenders.}  The average detection rate drops to 62.2\%, i.e., the adversarial images have better chances to escape human auditors. Detection rates decrease in every attack except SQ, while the median task completion time decreases to 66 seconds. The most and least stealthy attacks remain the same as Exp 1: WN with a 17.8\% detection rate and SQ with a 99.1\% detection rate. Noise-based evasion attacks are the most stealthy with a mean detection rate of 49.4\% (12.6\% decrease from Exp 1). The patch-based backdoors are the least stealthy with a detection rate of 89.0\% (3.7\% decrease). 

\vspace{1mm}\noindent\textbf{Exp III. Referenceless Defenders.} The average detection rate increases slightly from 62.2\% in Exp II to 67.9\% in Exp III. The distribution demonstrates a high correlation with Exp II with a Spearman correlation coefficient of 0.97. WN and SQ remain the most and least stealthy attacks with detection rates of 32.6\% and 98.6\%, respectively. Global transformation attacks become the most stealthy category with a 53.7\% mean detection rate. 

\vspace{1mm}\noindent\textbf{Observations on Attack Categories.} For attacks with different categories of injected adversarial perturbations, we observed distinct behaviors across three experiments: (1) The global-noise-based evasion attacks are consistently among the more stealthy types of attacks across all three experiments, with mean detection rates in the range of [$49\%, 61\%$]. While the perturbation budget introduces a limit to the per-pixel modifications, the injected noise pattern is still noticeable to 50\%+ of the users, with or without reference/knowledge to the benign dataset. (2) The global-noise-based backdoors are significantly less stealthy than the evasion attacks although they exploit similar types of perturbations. Such backdoor attacks attempt to use weak global noise to generate salient features to create robust backdoors, which appears to be difficult--the noise levels are higher than the noise-based evasion attacks, therefore, the backdoor attacks are noticeable to $70\%$ to $80\%$ of the users. (3) Local-patch-based backdoors are consistently the least stealthy attacks in all settings. The mean detection rates of 89\% to 92\% indicate that they can hardly escape human auditors. (4) Compared with other attack categories, global transformation attacks demonstrate a very unique pattern. They have a mean detection rate of 72.9\% in side-by-side comparisons in Exp I, while the mean detection rate significantly drops to $\sim$50\% in Exps II and III. In particular, the Instagram attack was detected at 98.6\% in Exp 1 but only 33\% to 39\% in Exps II and III. That is, users are significantly more likely to be fooled by such attacks without side-by-side comparisons. 

\begin{figure}[t]
\centering
	\includegraphics[width=0.98\linewidth]{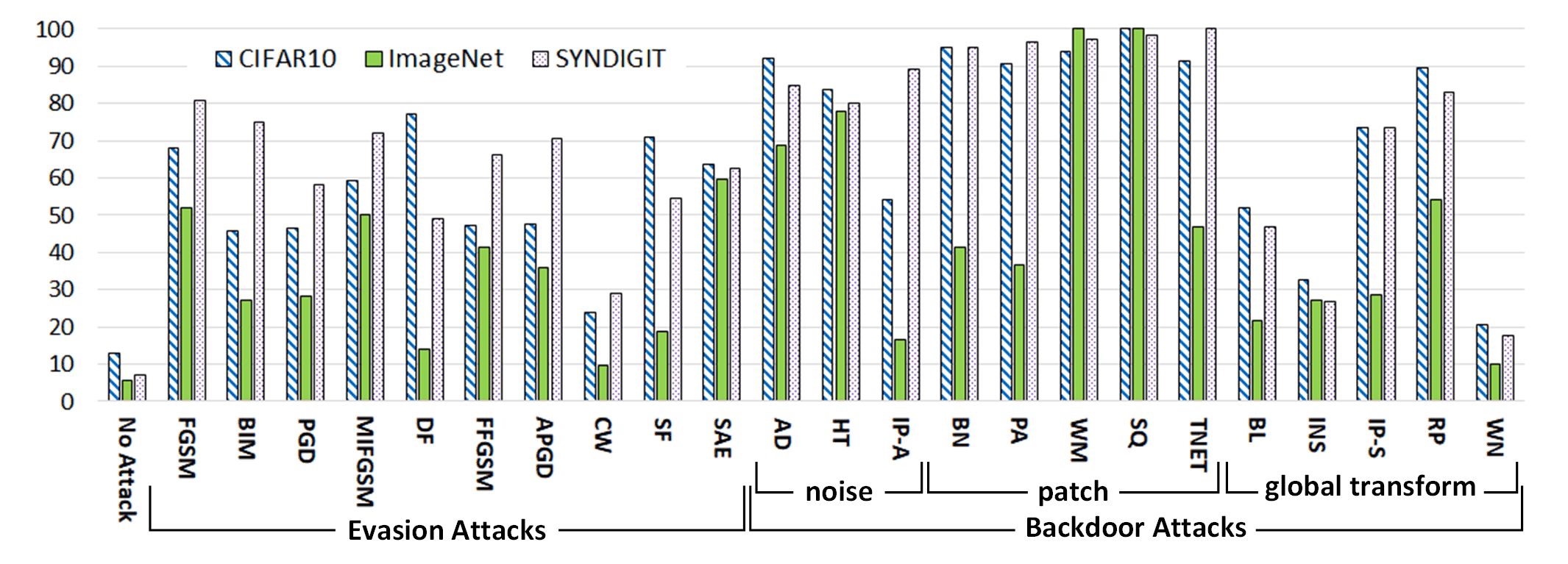}\vspace{-2mm}
	\caption{Detection rates (\%) for different datasets in Exp II: Knowledgeable Defenders. \vspace{-3mm}}
	\label{fig:dr-dataset}
\end{figure}


\vspace{1mm}\noindent\textbf{The Impact of Image Size and Content.} 
While we have observed highly consistent results across six datasets in numerical assessments, the user study results are different--our results show that the size and quality of the images may significantly affect user-perceived attack stealthiness. For instance, we present the Exp II detection rates for three datasets (CIFAR, ImageNet, SYNDIGIT) in Fig.  \ref{fig:dr-dataset}. We can observe the following: (1) For the auditors, smaller static patches (BN, PA, TNET) are stealthier in larger images (ImageNet). (2) Global transformation attacks (especially BL and WN) also appear to be stealthier in larger images since the per-pixel modification is weaker. (3) For images with more complex content (ImageNet), the noise-based evasion and backdoor attacks become stealthier to users/auditors. (4) For images with a relatively simple foreground and a clean background (e.g., SYNDIGIT), the detection rate is high in the majority of the attacks, i.e., regardless of the type of the added perturbation, the adversarial samples appear highly suspicious to human eyes.

\vspace{1mm}\noindent\textbf{Referenceless Defenders.} We initially expected the referenceless defenders to be the weakest. However, results in Section~\ref{sec:res} show that, the true detection rate increases in Exp III in comparison with Exp II. That is, the defenders/auditors tend to get more aggressive in identifying potentially adversarial samples, when they do not have reference to benign data. While the false positive rate increases, i.e., benign samples are mislabeled as suspicious, the true adversarial samples are more likely to be identified. In the literature, we have seen discussions that ``adversarial examples are likely to escape novice defenders when they are unfamiliar with the new testing samples or have not previously seen the testing dataset.'' Such claims are rejected by our experimental results. 

\vspace{1mm}\noindent\textbf{Error Rates with Benign Samples.}  In each experiment, a portion of users flagged the benign images as ``different'' or ``malicious''. The error rates are 5.23\%, 12.46\%, and 23.70\% in Experiments I, II, and III, respectively. The errors are explained by two main reasons: (1) there are always human errors in any type of data labeling, e.g., \cite{northcutt2021pervasive} reports a 3.3\% error rate across 10 image datasets and a 6\% error rate in ImageNet validation set. The 5.23\% base error rate in Exp I is in line with reports in the literature. (2) Some benign benchmarking images do appear to be suspicious, e.g., a significant portion of the tiny images contain blurs, unnatural edges, or objects. As shown in Fig. \ref{fig:dataset} (D), MNISTM is the most ``naturally suspicious'' dataset, with 30.3\% and 48.8\% error rates in Exp II and III, respectively. The 18.5\%  difference indicates that a small number of reference examples in Exp II effectively help the user/auditor to reduce benign errors. Meanwhile, the error rates with ImageNet remain under 5.5\% across all experiments, which shows that regular users are highly capable of recognizing benign high-resolution images even with limited or no reference.

\vspace{1mm}\noindent\textbf{MTurk Study.} We initially planned to utilize Amazon Mechanical Turk to conduct the same set of experiments. In the test with Exp I, we injected two pairs of digitally identical images and two pairs of totally different images in each questionnaire for quality control. With 200 questionnaires (all in Exp I, \$0.50 per questionnaire, workers with >95\% approval rates), we received almost completely random annotations. Even with the quality control questions, the detection rates were close to 50\% for both identical and totally different image pairs. This is consistent with recent reports on substantial decreases in MTurk data quality caused by fraudulent workers or bots \cite{bai2018evidence,chmielewski2020mturk,dreyfuss2018bot}. We eventually gave up the MTurk study due to the cost and data quality concerns. 

{
\begin{table}[t]
\caption{Correlation between numerical assessments and detection rates from the user study. Bold: top 3 correlations in each setting; red: p-value>0.05.}\label{tbl:correlation}
\vspace{-2mm}
\begin{scriptsize}
\begin{center}
\begin{tabular}{
p{11pt} <{\centering}
p{22pt} <{\centering}
p{22pt} <{\centering}
p{25pt} <{\centering}
p{22pt} <{\centering}
p{22pt} <{\centering}
p{24pt} <{\centering}
p{24pt} <{\centering}
p{24pt} <{\centering}
p{24pt} <{\centering}
p{25pt} <{\centering}
p{29pt} <{\centering}
p{29pt} <{\centering}
p{24pt} <{\centering}
}
\hline\hline
\multicolumn{13}{c}{\textbf{Exp I. Informed Defenders in All Attacks}}\\\hline
 & ${l}^{0}$ & ${l}^{1}$ & ${l}^{2}$ & ${l}^{\infty}$ & MSE & PSNR & SSIM & LPIPS & aHash & dHash & pHash & wHash \\
S & \textcolor{red}{-0.16} & 0.49 & \textbf{0.80} & 0.68 & 0.79 & \textbf{-0.82} & -0.77 & 0.78 & 0.67 & 0.62 & 0.75 & 0.66 \\

K & \textcolor{red}{-0.07} & 0.39 & \textbf{0.65} & 0.51 & 0.63 & \textbf{-0.67} & -0.62 & 0.63 & 0.54 & 0.48 & 0.58 & 0.52\\
\hline
 & FDL & FFL & ERGAS & SAM & {\tiny MSSSIM} & FSIM & VIF & UQI & Elastic & RE & ${l}^{1}$Cos & ${l}^{1}$ Clip \\
S & 0.61 & 0.78 & 0.68 & \textbf{0.82} & -0.76 & \textcolor{red}{-0.27} & \textcolor{red}{-0.37} & -0.77 & {0.79} & \textcolor{red}{0.35} & 0.49 & \textcolor{red}{0.04}\\

K & 0.49 & 0.62 & 0.53 & \textbf{0.67} & -0.61 & \textcolor{red}{-0.22} & \textcolor{red}{-0.29} & -0.61 & 0.63 & \textcolor{red}{0.31} & 0.39 & \textcolor{red}{0.12}\\

\hline\hline
\multicolumn{13}{c}{\textbf{Exp II. Knowledgeable Defenders in All Attacks}}\\
\hline
 & ${l}^{0}$ & ${l}^{1}$ & ${l}^{2}$ & ${l}^{\infty}$ & MSE & PSNR & SSIM & LPIPS & aHash & dHash & pHash & wHash \\
S & \textcolor{red}{-0.35} & \textcolor{red}{0.22} & 0.57 & 0.54 & 0.55 & \textbf{-0.59} & -0.58 & \textbf{0.64} & \textcolor{red}{0.40} & \textcolor{red}{0.32} & 0.53 & \textcolor{red}{0.41} \\

K & \textcolor{red}{-0.20} & \textcolor{red}{0.17} & 0.44 & 0.42 & 0.43 & \textbf{-0.47} & -0.45 & \textbf{0.50} & \textcolor{red}{0.29} & \textcolor{red}{0.22} & 0.40 & \textcolor{red}{0.32}\\
\hline
 & FDL & FFL & ERGAS & SAM & {\tiny MSSSIM} & FSIM & VIF & UQI & Elastic & RE & ${l}^{1}$ Cos & ${l}^{1}$ Clip \\
S & \textcolor{red}{0.35} & 0.55 & \textcolor{red}{0.45} & \textbf{0.63} & \textcolor{red}{-0.55} & \textcolor{red}{-0.03} & \textcolor{red}{-0.27} & \textcolor{red}{-0.52} & 0.55 & \textcolor{red}{0.20} & \textcolor{red}{0.22} & \textcolor{red}{-0.21} \\

K & \textcolor{red}{0.28} & 0.43 & \textcolor{red}{0.33} & \textbf{0.51} & -0.42 & \textcolor{red}{-0.06} & \textcolor{red}{-0.25} & -0.41 & 0.43 & \textcolor{red}{0.19} & \textcolor{red}{0.17} & \textcolor{red}{-0.06}\\\hline\hline

\multicolumn{13}{c}{\textbf{Exp III. Referenceless Defenders in All Attacks}}\\
\hline
 & ${l}^{0}$ & ${l}^{1}$ & ${l}^{2}$ & ${l}^{\infty}$ & MSE & PSNR & SSIM & LPIPS & aHash & dHash & pHash & wHash \\
S & \textcolor{red}{-0.30} & \textcolor{red}{0.14} & \textcolor{red}{0.45} & \textcolor{red}{0.38} & \textcolor{red}{0.42} & \textcolor{red}{-0.48} & \textcolor{red}{-0.50} & \textbf{0.59} & \textcolor{red}{0.30} & \textcolor{red}{0.23} & 0.47 & \textcolor{red}{0.31} \\

K & \textcolor{red}{-0.17} & \textcolor{red}{0.11} & 0.38 & \textcolor{red}{0.27} & 0.37 & -0.40 & -0.40 & \textbf{0.46} & \textcolor{red}{0.24} & \textcolor{red}{0.17} & 0.34 & \textcolor{red}{0.25} \\
\hline
 & FDL & FFL & ERGAS & SAM & {\tiny MSSSIM} & FSIM & VIF & UQI & Elastic & RE & ${l}^{1}$ Cos & ${l}^{1}$ Clip \\
S & \textcolor{red}{0.29} & \textcolor{red}{0.39} & \textcolor{red}{0.34} & \textbf{0.55} & \textcolor{red}{-0.45} & \textcolor{red}{-0.10} & \textcolor{red}{-0.28} & \textcolor{red}{-0.39} & \textcolor{red}{0.43} & \textcolor{red}{0.25} & \textcolor{red}{0.15} & \textcolor{red}{-0.24}\\

K & \textcolor{red}{0.23} & \textcolor{red}{0.34} & \textcolor{red}{0.25} & \textbf{0.44} & \textcolor{red}{-0.37} & \textcolor{red}{-0.10} & \textcolor{red}{-0.25} & \textcolor{red}{-0.31} & 0.37 & \textcolor{red}{0.23} & \textcolor{red}{0.12} & \textcolor{red}{-0.10}\\\hline\hline
\hline
\multicolumn{13}{c}{\textbf{Subset of Exp II. Knowledgeable Defenders: Noise-based Attacks.}}\\
\hline
 & ${l}^{0}$ & ${l}^{1}$ & ${l}^{2}$ & ${l}^{\infty}$ & MSE & PSNR & SSIM & LPIPS & aHash & dHash & pHash & wHash \\
S & 0.74 & 0.78 & 0.75 & 0.05 & 0.72 & -0.80 & -0.77 & \textbf{0.81} & \textcolor{red}{0.42} & \textcolor{red}{0.44} & \textcolor{red}{0.53} & 0.60\\

K & 0.56 & 0.67 & 0.63 & 0.06 & 0.60 & -0.67 & -0.65 & \textbf{0.70} & \textcolor{red}{0.33} & \textcolor{red}{0.32} & \textcolor{red}{0.43} & 0.49\\
\hline
 & FDL & FFL & ERGAS & SAM & {\tiny MSSSIM} & FSIM & VIF & UQI & Elastic & RE & ${l}^{1}$ Cos & ${l}^{1}$ Clip \\
S  & \textbf{0.81} & 0.71 & 0.70 & 0.79 & -0.64 & -0.78 & -0.75 & -0.68 & 0.73 & \textbf{0.82} & 0.78 & \textbf{0.84} \\

K & 0.69 & 0.59 & 0.57 & 0.67 & -0.54 & -0.64 & -0.64 & -0.56 & 0.61 & \textbf{0.72} & 0.67 & \textbf{0.72}\\\hline\hline

\multicolumn{13}{c}{\textbf{Subset of Exp III. Referenceless Defenders: Noise-based Attacks. }}\\
\hline
 & ${l}^{0}$ & ${l}^{1}$ & ${l}^{2}$ & ${l}^{\infty}$ & MSE & PSNR & SSIM & LPIPS & aHash & dHash & pHash & wHash \\
S & 0.60 & 0.71 & 0.70 & \textcolor{red}{0.04} & 0.66 & -0.75 & -0.70 & 0.75 & \textcolor{red}{0.37} & \textcolor{red}{0.37} & \textcolor{red}{0.41} & \textcolor{red}{0.47} \\

K & 0.46 & 0.60 & 0.61 & \textcolor{red}{0.05} & 0.59 & -0.64 & -0.60 & 0.65 & \textcolor{red}{0.33} & \textcolor{red}{0.30} & \textcolor{red}{0.36} & \textcolor{red}{0.40}  \\
\hline
 & FDL & FFL & ERGAS & SAM & {\tiny MSSSIM} & FSIM & VIF & UQI & Elastic & RE & ${l}^{1}$ Cos & ${l}^{1}$ Clip \\
 
S & \textbf{0.79} & 0.63 & \textcolor{red}{0.61} & 0.73 & \textcolor{red}{-0.54} & -0.75 & -0.69 & \textcolor{red}{-0.59} & {0.67} & \textbf{0.80} & 0.71 & \textbf{0.79}\\

K & \textbf{0.67} & 0.57 & \textcolor{red}{0.50} & 0.63 & \textcolor{red}{-0.48} & -0.61 & -0.59 & \textcolor{red}{-0.49} & 0.60 & \textbf{0.68} & 0.60 & \textbf{0.65}\\\hline\hline

\end{tabular}
\end{center}
\end{scriptsize}
\vspace{-5mm}
\end{table}
}

\subsection{Correlation: Numerical Metrics vs. User Perceived Stealthiness} \label{sec:cor}

Some of the image similarity/quality metrics in Sec. \ref{sec:num}) are designed to reflect the human visualization system. Now, we further investigate whether they could be used to assess the user-perceived attack stealthiness. To investigate the correlations between numerical analysis and the user studies, we employ two metrics: Spearman Rank Order Correlation Coefficient (S or SROCC) and  Kendall Rank Order Correlation Coefficient (K or KROCC). We do not use the popular Pearson correlation because our data do not follow normal distribution.

First, we report the correlation between the numerically assessed stealthiness (24 metrics in Tables \ref{tbl:evasion} and \ref{tbl:backdoor}) and user-perceived stealthiness (three user studies in Section \ref{sec:user}) across all 23 attacks. The results are shown in the top 3 sections in Table \ref{tbl:correlation}. We reject the null hypothesis when the p-value exceeds 0.05 and mark the correlation coefficients as red in the table. In Exp I, most of the image similarity/quality metrics fit well with the detection rates (DR): 11 out of 24 metrics show strong correlations in SROCC ($|S|{>}0.7$). That is, attacks assessed with lower distance/error/noise (or higher similarity/quality) are less likely to be identified in Exp I. In particular, SAM, PNSR, and ${l}^{2}$ demonstrate the strongest correlation with >0.8 on SROCC and >0.65 on KROCC.

However, the correlations are significantly weaker in Exps II and III. None of the metrics shows strong correlations with DR, while only 10 metrics show moderate correlations ($0.4{<}|S|{<}0.7$) with DR in Exp II in both SROCC and KROCC, and 2 metrics show moderate correlations in Exp III. Overall, SAM \cite{SAM} and LPIPS \cite{LPIPS} appear to be the most consistent with human perceptions in all three experiments, however, the correlations are only moderate with knowledgeable and referenceless defenders. Meanwhile, image similarity/quality metrics using structural information or deep features demonstrate better performance. 

Meanwhile, we also evaluate the correlation between the numerically assessed and user-perceived stealthiness of each category of attacks. We found strong correlations in global-noise-based attacks, as reported in the last two sections of Table \ref{tbl:correlation}. That is, the image similarity/quality metrics could properly evaluate the strength of the injected global perturbation in a way that is consistent with user perceptions. Last, the correlations are weak in other attack categories. 

In summary, none of the existing image similarity/quality metrics could  accurately assess users' perceptions of attack stealthiness in all attacks. However, some metrics have shown promising performance in a subset of the attacks. 

\section{The Stealthiness Assumption Revisited}\label{sec:dis}

Finally, we discuss the reflections of our findings w.r.t. ML attack stealthiness and practicality, which may contribute to a better understanding of the attacks. 

Most of the existing ML attacks in the literature explicitly or implicitly employ the stealthiness assumption. The underlying rationale is that it may be difficult for less experienced users to examine the DL model architecture, code, or parameters, however, even novice users could examine training/testing images and identify anomalies. Here we revisit such assumptions and discuss their practicality based on our findings. 

\noindent$\bullet~$\textbf{Evasion attacks.} Many evasion attacks in the literature explicitly claim to be stealthy and enforce stealthiness using a perturbation budget. Results from our user study show that majority of them are still highly noticeable even to users without reference images. Meanwhile, we also show that some image similarity/quality metrics could be employed to effectively predict the user-perceived stealthiness of the attack images. For an attack to be stealthy in practice, a tight threshold (high similarity or low distance) must be set for the metrics. 

\noindent$\bullet~$\textbf{Data poisoning backdoors.} Conventional data poisoning backdoors do not assume stealthiness for the poisoned training samples, since they are designed to carry labels that are visually wrong. Clean label poisoning attacks have been proposed to tackle this issue by using adversarial samples that appear to carry the correct labels. However, as demonstrated in our user study, the clean label poisoning attacks are highly detectable by users with or without reference. When the attack attempts to train the DNN to learn a \textit{salient} feature from  weak perturbations, the actual perturbation is too strong to escape human eyes.  

\noindent$\bullet~$\textbf{Patch-based backdoors.} In general, the patch-based backdoors appear to be the least stealthy, except for the tiny patches on larger images, e.g., BN and TNET on ImageNet. In order for the patches to be learned by the victim DNN as robust features, the patches must have reasonable size and salient visual features, which is against the stealthiness assumption. 

Finally, we would like to answer this question from our experiments and observations: \textit{what makes an attack stealthy, i.e., what makes the adversarial image less likely to be identified by a human auditor?} 

\vspace{1mm}\noindent\textbf{(1) Scale of Perturbation Matters. } Attacks with \textit{extremely low} perturbation budget are more likely to escape human evaluators, especially in global-noise-based attacks. In practice, a global perturbation budget of 8, or a normalized $l^1$ in the range of [5, 8] appears to be too strong to be unnoticeable. 

\noindent\textbf{(2) Size of Perturbed Region Matters.} Attacks that modify a relatively \textit{smaller} portion of the victim image, i.e., attacks with a very small $l^0$, are more likely to escape evaluators. 

\noindent\textbf{(3) Image Content Matters.} Attacks on images that are fully filled with foreground and have complex content are more likely to escape human evaluators, while most of the attacks, especially the noise-based ones, are very noticeable on a clean background (e.g., blue sky, white wall). 

In practice, it would be very difficult to design an attack that satisfies both (1) and (2), since such weak perturbations are unlikely to trigger robust responses from the victim DNN.  We have not seen such an attempt in the literature yet. Meanwhile, when an attack achieves one of (1) and (2) on images satisfying (3), the attack is quite stealthy to human eyes, e.g., CW on ImageNet. Last, a different design philosophy has been proposed in the literature, e.g., SAE and INS. Such attacks do not seek to minimize the perturbation or to hide the perturbation, instead, they attempt to apply special visual effects on the victim images, so that the adversarial samples, although significantly different from the victim images, appear to be benign by themselves. The adversarial samples can hardly escape side-by-side comparisons, however, they could better fool the auditors if they do not have reference to the benign dataset. 

In summary, we argue that user-based evaluation is the golden standard to validate the stealthiness assumptions in adversarial ML. For global-noise-based attacks, some image quality metrics could be employed to predict attack stealthiness in lieu of a user study. In such metrics, only attacks with extremely low perturbation (high similarity/quality) may result in practically stealthy attacks. Very few attacks in the literature have achieved this goal.

\section{Conclusion}\label{sec:con}
In this paper, we present the first large-scale comparative experimental study of the stealthiness of evasion and backdoor attacks against deep learning systems. We have implemented 20 attacks (23 different settings) on six benchmarking datasets. We first present numerical measurements using 24 image quality metrics on all the attacks. Next, we design a user study of three questionnaires that ask users to identify potentially adversarial images in three different settings. With 1,500+ responses and 30,000+ labeled images, we find that majority of the attacks in the literature are not really stealthy to human eyes. We also identify the factors that impact attack stealthiness, e.g., the type of perturbation, the size, quality, and content of the victim images. We further examine the correlations between numerically assessed image similarity/quality and user-perceived stealthiness and re-visit the stealthiness of the attacks with our findings.  

\section*{Acknowledgements}

Zeyan Liu, Fengjun Li and Bo Luo were sponsored in part by NSF awards IIS-2014552, DGE-1565570, DGE-1922649, and the Ripple University Blockchain Research Initiative. Zhu Li was supported in part by NSF award 1747751. The authors would like to thank the anonymous reviewers for their valuable comments and suggestions. The authors would like to thank all the participants of the user studies. 

\bibliographystyle{splncs04}
\bibliography{ref}

\begin{subappendices}

\end{subappendices}

\end{document}